\documentclass[aps,prl,groupedaddress, twocolumn,nofootinbib,superscriptaddress]{revtex4-1}
\newcommand{\be}{\begin{eqnarray}}
\newcommand{\ee}{\end{eqnarray}}
\usepackage{graphicx,epsfig}
\usepackage{amsmath}
\usepackage{amsfonts}
\usepackage{fancyhdr}
\usepackage{amsmath}
\usepackage{mathrsfs}
\begin{document}


\title{Beyond dimensional analysis: Higgs and new Higgs inflations {\it do not} violate unitarity}

\author{Albert Escriv\`a}
\email{albert.escriva@fqa.ub.edu}
\affiliation{Institut de Ci\`encies del Cosmos, Universitat de Barcelona, Mart\'i i Franqu\`es 1, 08028 Barcelona, Spain}
\affiliation{Departament de F\'isica Qu\`antica i Astrof\'isica, Facultat de F\'isica, Universitat de Barcelona, Mart\'i i Franqu\`es 1, 08028 Barcelona, Spain}
\author{Cristiano Germani}
\email{germani@icc.ub.edu}
\affiliation{Institut de Ci\`encies del Cosmos, Universitat de Barcelona, Mart\'i i Franqu\`es 1, 08028 Barcelona, Spain}


\begin{abstract}
Na\"ive dimensional analysis seems to suggest possible unitarity violations in the framework of the Higgs and new Higgs inflationary scenarios. These violations seem to happen around the value in which the potential energy, per given Higgs boson's vacuum expectation value, crosses the perturbative cut-off scale calculated around the electroweak vacuum. Conversely to these expectations, in this paper we show that, by using an exact analysis of the background dependent cut-off scale, and by including the contribution of the phase-space volume in the perturbative scattering amplitudes of scalars, no violation of (perturbative) unitarity might ever happen during the whole Universe evolution. 
\end{abstract}


\maketitle

\section{The na\"ive story}
Because inflation is supposed to generate a homogeneous and isotropic Universe (at the background level), a natural Inflaton candidate is a spin 0 particle. The standard model of particle physics (SM) contains already a scalar field: the Higgs boson. Therefore one may be tempted to consider the very economical scenario in which the Higgs boson is also the inflaton (this is a very old story, see for example \cite{salopek}). 
However, as it is well known, the Higgs boson has a too steep potential to generate a successful inflation. Assuming no extra degrees of freedom in Nature, rather than the ones of the standard model of particle physics and gravity, at least up to the inflationary scale, lead to only two possibilities. The first one is that the actual Higgs potential becomes shallower after certain scales (Higgs inflation) \cite{salopek} (for this scenario with current precision data see \cite{higgs1,*higgs2}), and the other is that the gravitational friction acting on the Higgs boson is stronger than the one already provided by standard General Relativity (GR) (new Higgs inflation) \cite{new}. After this is achieved at tree-level, loops analysis can be done such to consider the contribution of the running coupling constants to inflation (see \cite{higgs1} for Higgs inflation and \cite{loop-new-higgs} for new Higgs inflation).

However, it has already been criticised the use of these scenario on the base of a putative perturbative unitarity violation (for the latest rebuttal on this see \cite{kehagias-unitarity}). The story goes along those lines \footnote{Strictly speaking this has been extensively analysed only on the Higgs inflation of \cite{salopek}. However, the same qualitative analysis can be done in the new Higgs inflation \cite{new}.}: Although the Higgs inflationary scenarios make only use of the SM degrees of freedom and gravity, they are non-renormalizable effective field theories (EFT). The Higgs boson and/or the graviton are no canonical degrees of freedom in these EFTs. Therefore, in order to figure out the perturbative unitarity violation scale, one has to diagonalise the graviton-Higgs system and then, check the scales ($V_m$) suppressing the non-renormalizable operators constructed with $m$ powers of the diagonalised field fluctuations. This procedure is however background dependent and so the vertices $V_m$. 

The diagonalization of the Higgs-gravity system does not have an analytical form. For this reason, the usual approach has been to consider the perturbative cut-offs ($\Lambda_m$) constructed on the vertices $V_m$, only around the electroweak ($\phi=v$) and large Higgs vacuum expectation values (vevs) ($\phi=\phi_\ell$) scales. 

Within the above approximation, one finds that the na\"ive cut-off in dimensional analysis, i.e. for $\Lambda_m\sim V_m$, is obtained in the $m\rightarrow\infty$ limit. In particular, since cut-off are too background dependent, one readily finds $\Lambda_{v}< \Lambda_{\ell}$. 

As the potential energy of the Higgs boson becomes larger than $\Lambda_{v}$ for $\phi\ll \phi_\ell$, extrapolating the behaviour of the cut-off in the transition region $v<\phi<\phi_\ell$, standard lore was to advocate a perturbative unitarity violation there.

The ``reason" is the following: the fact that the background energy is larger than the perturbative cut-off around the background chosen, does not mean by itself that unitarity is violated. However, quite conservatively, one might consider that the potential energy of the background represents a reservoir energy available for random scatterings between scalar fluctuations. In this case, there can be a non-vanishing probability that a scattering with momenta of order the background energy happens, so violating unitarity. 
Obviously, this conclusion may be very superficial as does not really follow from a thorough examination of high energy scatterings in this system (see for example \cite{he1,*he2, bezsib}). 

In any cases, even blindly using the argument that the potential energy density sets the maximal momenta for random scatterings, we show that the conclusion of unitarity violation of these systems is premature. The main reason is that the exact background cut-offs, calculated by considering the unitarity bounds of tree-level scattering amplitudes, instead of na\"ive dimensional analysis of previous literature, show unequivocally that the potential energy is always way below the perturbative unitarity violating scale. Specifically, for large $m$, $\Lambda_m\gg V_m$ instead of the dimensional analysis $\Lambda_m\sim V_m$. We will prove that by considering $2\rightarrow n$ scatterings.

Although the question of whether the potential of Higgs inflationary scenarios is fine tuned or not still remain, our results show that Higgs and new Higgs inflationary scenarios are robust EFT during the whole Universe evolution.
 
 \section{Higgs inflation}
 In the Higgs inflationary scenario the Higgs boson's ($\phi$) action, at large field values (in the unitary gauge, neglecting SM couplings and the Higgs mass) is
 \be
 \!\!\!\!\!\!\!S_{H}=\frac{1}{2}\int d^4x \sqrt{-\tilde g}\Big[M_p^2\left(1+\frac{\xi\phi^2}{M_p^2}\right)\tilde R-\tilde\partial_\alpha\phi\tilde\partial^\alpha\phi
 -\frac{\lambda}{2}\phi^4\Big]\ \nonumber,
 \ee 
 where $M_p=2.435\cdot 10^{18}\ {\rm Gev}$ is the Planck mass, $\tilde{(\cdot)}$ means evaluated with the metric $\tilde g$, $\lambda\simeq 0.13$ is the tree-level Higgs quartic coupling and finally $\xi\gg 1$ is a constant defining a new mass scale $M=M_p \xi^{-1/2}$. Note that the conformal term multiplying the Ricci scalar $R$ is not unique as we could have chosen any function of $\phi$. We will however follow \cite{salopek} for our analysis. 
 
Although it might seem that in the above action the Higgs boson is non-minimally coupled to gravity it is not. The reason is that in the canonical system, the so-called "Einstein frame", the field $\phi$ is minimally coupled.
 
By making the conformal re-scaling $g_{\alpha\beta}=\Omega^2 \tilde g_{\alpha\beta}$, with $\Omega^2=\left(1+\frac{\xi\phi^2}{M_p^2}\right)$ we will end up with the canonical action
\be
 S_{H}=\frac{1}{2}\int d^4x \sqrt{-g}\Big[M_p^2 R-\partial_\alpha\chi\partial^\alpha\chi- 2 U(\chi)\Big]\ ,
 \ee 
 where the relation between the new canonical variable $\chi$, the Higgs boson and the new potential $U$ are
 \be
 \frac{d\chi}{d\phi}&=&\sqrt{\frac{\Omega^2+6\xi^2\phi^2/M_p^2}{\Omega^4}}\ ,\cr
 U(\chi)&=&\frac{\lambda}{4}\frac{\phi(\chi)^4}{\left(1+\frac{\xi \phi(\chi)^2}{M_p^2}\right)^2}\ .
 \ee
 After this field redefinition, we see that the non-renormalizability of the action is codified by a non-renormalizable potential (and of course by gravity). It is easy to see that for small values of the Higgs vacuum expectation value $\phi\ll M/\sqrt{\xi}$, the potential is well approximated by the standard electroweak potential and the Higgs boson is approximately canonical, while for large vevs $\phi\gg M$ the potential becomes exponentially flat so to accommodate inflation.
 
From the perspective of perturbative analysis as seen from the electroweak vacuum, i.e. expanding the Higgs boson potential around the $\chi=0$ vev, perturbative unitarity is violated at the energy $M/\sqrt{\xi}$. Therefore, one may ask the question of whether perturbative unitarity is violated somewhere before reaching the energy scales interesting for inflation. A second question would then be whether or not the modified Higgs potential is fine tuned at one or more loops level. Here, we will only focus on the question of whether perturbative unitarity is violated and leave the question of possible fine tunings for future work.

In principle, the fact that the Higgs vev is large, implying a ``large" energy density, does not necessarily imply a violation of unitarity, as we have already discussed. However, one might argue that the energy of the background should always be below the perturbative cut-off energy scale. We will not discuss here whether this assumption is correct or not, what we will show is that, even accepting this conservative prejudice, neither the Higgs nor the new Higgs inflationary scenarios violate perturbative unitarity during the whole Universe evolution. The reason is that in these scenarios, the unitarity violating scale is a background dependent quantity as we shall now discuss.

The procedure to obtain the background dependent cut-off is to expand the potential in some background $\chi = \chi_{0} + \delta \chi$ 
\begin{equation}
U(\chi+\chi_{0})=U(\chi_{0})+\sum_{m=1}^{\infty}\frac{1}{m!}\frac{d^m U(\chi)}{d \chi^m}\Big|_{\chi=\chi_0}(\delta \chi)^m\ ,
\end{equation}
and then calculate the cross-sections $\sigma$ of $\delta\chi$ particles due to any non-renormalizable vertices. In particular, without lost of generality, we will focus on the $2\rightarrow n$ scatterings with $n>2$. The unitarity bound is then 
\be
\sigma[2\rightarrow n]\leq \frac{4\pi}{s}\ ,
\ee
where $\sigma$ is the cross-section of the process.

The structure of these scattering amplitudes are very reminiscent of the multiple-particles scattering amplitudes of would-be Goldstone bosons in the electroweak theory, after symmetry breaking. There, dimensional analysis would suggest that the $2\rightarrow n$ scattering, at large $n$, would saturate the unitarity bound at the electroweak breaking scale $v$. However, kinematic conditions suggest at least a linear growing with the number of particles. This puzzle was resolved by \cite{factorial,*factor2} by noticing that dimensional analysis is grossly wrong for large $n$ due to the neglected phase-space volume contribution to the scattering amplitudes. Precisely the same fact that helped unitarity in the large $n$ scatterings in electroweak theory, will solve the unitarity issues in Higgs inflationary models. 

Here we will only focus on contact diagrams. Note that, in the $2\rightarrow n$ scattering amplitudes, also cascades of tree-level amplitudes with lower number of particles contribute. There, the sum of those diagrams could violate unitarity. However, if every single vertex does not, by doing a careful (Borel) summation, the final scattering amplitude {\it should} not violate unitarity too. Nevertheless, this is an old issue that still afflicts even renormalizable theories, e.g. the standard electroweak $\phi^4$ theory. By assuming that this computational matter is solved in quantum field theory, we will not further investigate it here. Indeed, this issue is way beyond the scope of the present paper which aims to show that contact diagrams do not violate unitarity, as conversely claimed in previous literature. In addition, as we are not aiming to find the precise value of the scattering amplitude of $2\rightarrow n$ scatterings, but rather just check for unitarity, we will not consider cascading diagrams.

Taking into account the symmetry factors due to the fact that we are considering a scattering of $m=n+2$ particles, the vertex involving $m$ particles is
\begin{equation}
V_{m}(\chi_0)=\left(\frac{d^m U(\chi_0)}{d \chi^m}\right)^{\frac{1}{4-m}}\ .
\end{equation}
The perturbative unitarity bound is then saturated at the energy (ignoring angular dependence)
\be
\Lambda_m(\chi_0)={\cal O}(1)\Big|V_m(\chi_0)\Big| F_m^{\frac{1}{2(m-2)}}\ ,
\ee 
where the factor
\be
F_m=2^{4m-10}\pi^{2m-6}(m-3)!(m-4)!
\ee
comes form the phase-space integration \cite{factorial,*factor2}\footnote{Note that, conversely to \cite{factorial,*factor2} we are rewriting the scattering amplitude in function of $m$ and not $n$. In addition, since we are considering equivalent particles, the $\delta$ of \cite{factorial,*factor2} is $2$.}. 

Interestingly, $F_m^{\frac{1}{2(m-2)}}\rightarrow \frac{4\pi m}{e}$ for large $m$ making the perturbative cut-off linearly growing with the number of particles involved in the scattering. This fact was completely overlooked in the literature, for Higgs and new Higgs inflationary scenarios. That, lead to wrong claims about the perturbative unitarity violation in Higgs and new Higgs inflationary scenarios.

\begin{figure}
 
     \includegraphics[width=1\linewidth]{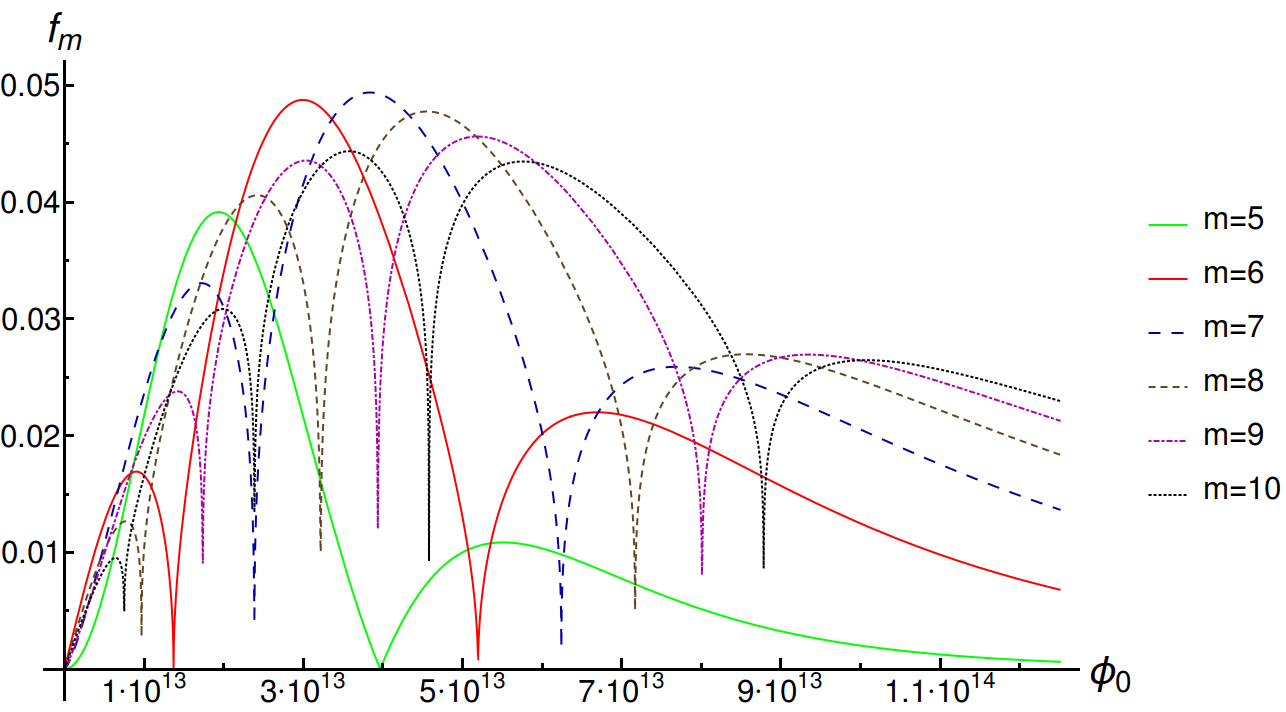} 
\caption{Numerical computation of the $f_{m}$s for Higgs inflation. The parameter $\xi\simeq 1.8\times 10^4$ is obtained from \cite{higgsspectra} by using $n_s\simeq 0.9655$ \cite{planck}. The picks corresponds to zero's of the $f_m$s. \label{fig1}}
\end{figure}

Forgetting the ${\cal O}(1)$ contribution to the perturbative cut-off \footnote{This is very conservative as the ${\cal O}(1)$ will help unitarity.}, we can now check whether $f_m(\chi_0)\equiv\frac{U(\chi_0)^{1/4}}{\Lambda_m(\chi_0)}$ ever exceeds one signalling a possible unitarity violation in the background chosen. As can be seen from the figure \ref{fig1}, there is a maximum of $f(\chi_0)$ close to where we would expect it, i.e. around the vev $\phi_c\sim M_p/\xi$ (the exact value depends on the vertex chosen). In the same figure we can see that the unitarity is never violated at least up to $m=10$. What is even more interesting is that for $m\geq 7$ the maximum values decreas and move to the right. Then the $m$-curve re-crosses the $(m-1)$s at values of $\phi$ increasing with $m$. However, this happens in the unitarity region. Unfortunately, due to the complexity of the function $\chi(\phi)$, $\Lambda_m$s with $m>10$ are exponentially expensive to calculate. 

Because of the shifting of the re-cross point towards large $\phi$ values, our intuition from the $m\leq 10$ curves is that there is always an $m_*(h)$ such that $\frac{f_{m_*}}{f_{m>m_*}}>1$, where $\frac{dm_*}{dh}>0$. If this is true then ${\rm Max}\left[f_m(\chi_0)\right]\sim f_7(\chi(\phi_c))<1$ and thus no unitarity violation may happen. 

To back up our intuition we will rely on some approximation that can be done in the innocuous region\footnote{I.e. in the region in which $f_m<1$ for all $m$.} $M_{p}/\xi \ll \phi\ll M_{p} / \sqrt{\xi}$. There, for large $m$, $V_{m} \approx \frac{\phi^2 \xi}{M_{p}} \left[\frac{\xi^6 \phi^6}{\lambda M_{p}^6}\right]^{1/(m-4)}$ \cite{bezsib}\footnote{In passing we noted that the computation of \cite{linde} is here wrong.}. It is then easy to check that, in this approximation and for the value of the parameter chosen in Fig.\ref{fig1}, $\frac{f_{m_*}}{f_{m>m_*}}>1$, with $m_*=34$, which backs up our expectations.

Concluding, we have shown that, within the Higgs inflationary model, no unitarity is violated in scalar scatterings. Note that the same conclusions was guessed by \cite{bezsib,linde}. However, as we have seen, the maximal peak of $f_m$ happens around $\phi_c\sim M_p/\sqrt{\xi}$. Precisely here, the approximated analysis of \cite{bezsib,linde} is invalid. In addition, our numerical results show that, if we had used $V_m$ instead of $\Lambda_m$ as in \cite{bezsib,linde}, we would have probably obtained a violation of unitarity.

\section{New Higgs inflation}

In the new Higgs inflation of \cite{new} the action is
 \be
 \!\!\!\!\!\!\!S_{nH}=\frac{1}{2}\int d^4x \sqrt{-\tilde g}\Big[M_p^2\tilde R-\left(\tilde g^{\alpha\beta}-\frac{\tilde G^{\alpha\beta}}{M^2}\right)\tilde\partial_\alpha\phi\tilde\partial_\beta\phi
 -2V(\phi)\Big]\ \nonumber,
 \ee 
 with $V(\phi)=\frac{\lambda}{4}\phi^4$. However, in this form, the previous analysis is not directly applicable.

By looking at the Hubble equations (see for example \cite{new}), on an homogeneous and isotropic background one easily find the bound $\dot\phi^2<\frac{2}{3}M^2 M_p^2$. In fact, in average $\langle\dot\phi^2\rangle\ll\frac{2}{3}M^2 M_p^2$ \cite{nina}. As discussed before, we are interested in the case in which some canonical scalar quanta could spontaneously scatter with up to the background energy. Using the same notation as used previously, we are therefore interested in the case in which $\partial_\alpha\delta\chi\partial^\alpha\delta\chi\leq U(\chi_0)$. As we shall soon see, the relation between $\chi$ and $\phi$ is $(\partial\chi)^2=(1+\frac{V(\phi)}{M^2 M_p^2})(\partial\phi)^2$, where $V(\phi)=\frac{1}{4}\lambda\phi^4$ and $U(\chi)\equiv V(\phi(\chi))$. Then, we immediately see that, in our region of interest, $(\partial\delta\phi)^2<M^2 M_p^2$, thus $\frac{\partial_\alpha\phi\partial_\beta\phi}{M^2 M_p^2}<1$. Although this might already be enough to solve unitarity issues in the new Higgs inflation \cite{yuki}, we will once again discuss the validity of this EFT in terms of scattering amplitudes. 

In this case we can use a disformal transformation of the metric \cite{parvin} $g_{\alpha\beta}=\tilde g_{\alpha\beta}+\frac{\partial_\alpha\phi\partial_\beta\phi}{M^2 M_p^2}$, to obtain
\be
 \!\!\!\!\!\!\!\!\!S_{nH}=\frac{1}{2}\int d^4x \sqrt{-g}\Big[M_p^2 R-\left(1+\frac{V(\phi)}{M^2M_p^2}\right)\partial_\alpha\phi\partial_\beta\phi
 -2V(\phi)+\cr
 +{\rm higher\ powers\ of\ derivatives}\Big]\ \nonumber,
\ee
where the operators with higher powers of derivatives will be suppressed by higher powers of $M^2 M_p^2$. Since we are interested in the leading contribution to the $m$-particles scatterings, and we do not aim for precision physics, we will neglect those.

\begin{figure}[h]
\includegraphics[width=0.50 \textwidth]{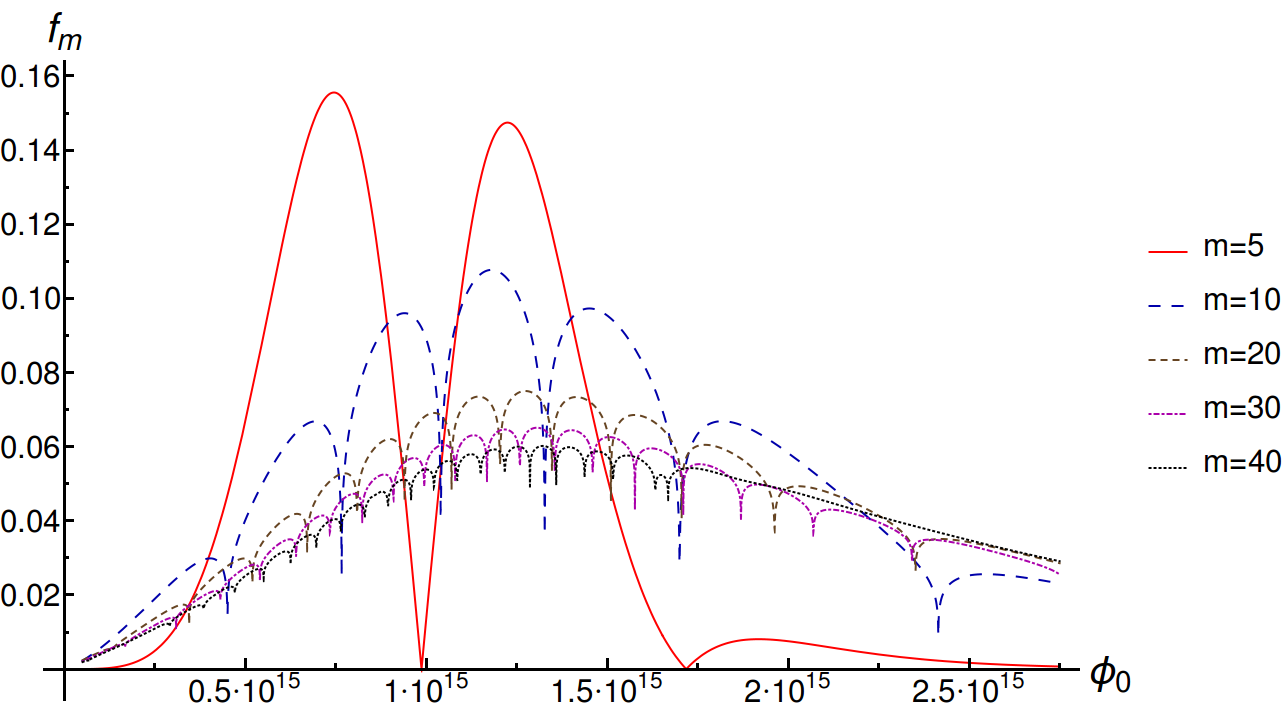}
\caption{Numerical computation of the $f_{m}$s for new-Higgs inflation. The parameter $M\simeq 7.2\times 10^{10}$GeV is obtained from \cite{newhiggsspectra} by using $n_s\simeq 0.9655$ \cite{planck}. As before, the picks corresponds to zero's of the $f_m$s.\label{fig2}}
\end{figure}

At this point, the discussion is immediately connected to the previous one with
\be
\frac{d\chi}{d\phi}=\sqrt{1+\frac{V(\phi)}{M^2M_p^2}}\ ,\cr
U(\chi)=V(\phi(\chi))\ .
\ee
Once the disformal transformation is performed, the good thing is that in the new Higgs inflation the computation of the vertices is not expensive, thus we can numerically plot the $f_m$s even for large $m$ values. In this case, as can be seen from the figure \ref{fig2}, $f_{m_1}>f_{m_2}$ for {\it any} $m_1>m_2$. Then, since unitarity is not violated for $f_5$, it is never. 

\section{Conclusion}
It has been a several years long debate, starting from \cite{espinosa, *burgess, *hertzberg} until \cite{kehagias-unitarity}, whether or not the Higgs inflation of \cite{shaposhnikov} violates tree-level unitarity. Lately, by using some non-perturbative techniques, the Authors of \cite{casadio}, studied loop corrections to the Higgs inflation and found indications that unitarity might not be violated, order by order. In this paper we show that this indeed exactly happens, i.e. Higgs inflation {\it does not} violate perturbative unitarity during the whole history of the Universe. 

The missing ingredients of previous analysis were basically two: the first one was simply computational. All analysis done before the current paper where based on extrapolations of the background dependent cut-offs behaviour on the region of the putative unitarity violation. The exact numerical calculation deviates from this. 

The second and crucial missing piece, was the actual calculation of the unitarity bound. All previous analysis were only performed by the use of dimensional analysis. In other words they were only based on checking whether the potential energy of the background would become larger than the suppression scales appearing in the expanded Higgs inflationary scenario, around the background chosen. There, typically, the violation of unitarity would appear in correspondence to the highest dimensional operator suppressed by a scale $V_m$, i.e. in a scattering of large number of ($m$) particles. 

However, in the latter case, the phase-space volume linearly grows with $m$ making the true cut-off $\Lambda_m$ also growing with $m$, i.e. $\Lambda_m\sim m V_m$. Contrary to the standard lore, we showed that larger and larger dimensional operators correspond to more and more unitary scatterings. This proved the Higgs inflation to be unitary during the whole Universe evolution.

We then analysed the new Higgs inflationary scenario of \cite{new}, where the Higgs boson is derivatively coupled to curvature. There, we have shown that, at least during the evolution of the Universe, a disformal transformation of the metric can be done so to find an approximate ``Einstein frame". In this case we were able to apply the same analysis performed in the Higgs inflationary case finding {\it no perturbative unitarity} violation for the whole history of the Universe.

A final comment is here due. In unitary gauge, the Higgs boson is a real scalar field $\phi$ coupled to gauge bosons. In Higgs inflation, at large field values (see \cite{bezsib}) the mass scale of non-abelian vectors is $M_V\simeq g M_p/\sqrt{\xi}$, where $g\ll 1$ is the non-abelian (strong) coupling. In this sector unitarity is then violated at a scale $\Lambda_V\sim M_V/g$. Because $\lambda\ll 1$, the cut-off scale in the vector scatterings is parametrically smaller than in the scalar sector, but still way larger than the scalar potential during inflation. Thus, no unitarity violation is expected there. 

In new Higgs inflation the situation is way better as, for large field values, the gauge bosons mass scale is parametrically {\it larger} than the scalar cut-off, see \cite{newhiggsspectra}.

\begin{acknowledgments}
CG is supported by the Ramon y Cajal program and partially supported by the Unidad de Excelencia Mar\'ia de Maeztu Grant No. MDM-2014-0369 and FPA2013-46570-C2-2-P grant. CG would like to thank 
Yukawa Institute for Theoretical Physics for hospitality and Alex Kehagias for comments. AE thanks useful discussions with Federico Mescia.
\end{acknowledgments}


\bibliography{refs.bib}

\begin{thebibliography}{23}%
\makeatletter
\providecommand \@ifxundefined [1]{%
 \@ifx{#1\undefined}
}%
\providecommand \@ifnum [1]{%
 \ifnum #1\expandafter \@firstoftwo
 \else \expandafter \@secondoftwo
 \fi
}%
\providecommand \@ifx [1]{%
 \ifx #1\expandafter \@firstoftwo
 \else \expandafter \@secondoftwo
 \fi
}%
\providecommand \natexlab [1]{#1}%
\providecommand \enquote  [1]{``#1''}%
\providecommand \bibnamefont  [1]{#1}%
\providecommand \bibfnamefont [1]{#1}%
\providecommand \citenamefont [1]{#1}%
\providecommand \href@noop [0]{\@secondoftwo}%
\providecommand \href [0]{\begingroup \@sanitize@url \@href}%
\providecommand \@href[1]{\@@startlink{#1}\@@href}%
\providecommand \@@href[1]{\endgroup#1\@@endlink}%
\providecommand \@sanitize@url [0]{\catcode `\\12\catcode `\$12\catcode
  `\&12\catcode `\#12\catcode `\^12\catcode `\_12\catcode `\%12\relax}%
\providecommand \@@startlink[1]{}%
\providecommand \@@endlink[0]{}%
\providecommand \url  [0]{\begingroup\@sanitize@url \@url }%
\providecommand \@url [1]{\endgroup\@href {#1}{\urlprefix }}%
\providecommand \urlprefix  [0]{URL }%
\providecommand \Eprint [0]{\href }%
\providecommand \doibase [0]{http://dx.doi.org/}%
\providecommand \selectlanguage [0]{\@gobble}%
\providecommand \bibinfo  [0]{\@secondoftwo}%
\providecommand \bibfield  [0]{\@secondoftwo}%
\providecommand \translation [1]{[#1]}%
\providecommand \BibitemOpen [0]{}%
\providecommand \bibitemStop [0]{}%
\providecommand \bibitemNoStop [0]{.\EOS\space}%
\providecommand \EOS [0]{\spacefactor3000\relax}%
\providecommand \BibitemShut  [1]{\csname bibitem#1\endcsname}%
\let\auto@bib@innerbib\@empty
\bibitem [{\citenamefont {Salopek}\ \emph {et~al.}(1989)\citenamefont
  {Salopek}, \citenamefont {Bond},\ and\ \citenamefont {Bardeen}}]{salopek}%
  \BibitemOpen
  \bibfield  {author} {\bibinfo {author} {\bibfnamefont {D.~S.}\ \bibnamefont
  {Salopek}}, \bibinfo {author} {\bibfnamefont {J.~R.}\ \bibnamefont {Bond}}, \
  and\ \bibinfo {author} {\bibfnamefont {J.~M.}\ \bibnamefont {Bardeen}},\
  }\href {\doibase 10.1103/PhysRevD.40.1753} {\bibfield  {journal} {\bibinfo
  {journal} {Phys. Rev.}\ }\textbf {\bibinfo {volume} {D40}},\ \bibinfo {pages}
  {1753} (\bibinfo {year} {1989})}\BibitemShut {NoStop}%
\bibitem [{\citenamefont {Bezrukov}\ \emph {et~al.}(2015)\citenamefont
  {Bezrukov}, \citenamefont {Rubio},\ and\ \citenamefont
  {Shaposhnikov}}]{higgs1}%
  \BibitemOpen
  \bibfield  {author} {\bibinfo {author} {\bibfnamefont {F.}~\bibnamefont
  {Bezrukov}}, \bibinfo {author} {\bibfnamefont {J.}~\bibnamefont {Rubio}}, \
  and\ \bibinfo {author} {\bibfnamefont {M.}~\bibnamefont {Shaposhnikov}},\
  }\href {\doibase 10.1103/PhysRevD.92.083512} {\bibfield  {journal} {\bibinfo
  {journal} {Phys. Rev.}\ }\textbf {\bibinfo {volume} {D92}},\ \bibinfo {pages}
  {083512} (\bibinfo {year} {2015})},\ \Eprint {http://arxiv.org/abs/1412.3811}
  {arXiv:1412.3811 [hep-ph]} \BibitemShut {NoStop}%
\bibitem [{\citenamefont {Bezrukov}\ and\ \citenamefont
  {Shaposhnikov}(2009)}]{higgs2}%
  \BibitemOpen
  \bibfield  {author} {\bibinfo {author} {\bibfnamefont {F.}~\bibnamefont
  {Bezrukov}}\ and\ \bibinfo {author} {\bibfnamefont {M.}~\bibnamefont
  {Shaposhnikov}},\ }\href {\doibase 10.1088/1126-6708/2009/07/089} {\bibfield
  {journal} {\bibinfo  {journal} {JHEP}\ }\textbf {\bibinfo {volume} {07}},\
  \bibinfo {pages} {089} (\bibinfo {year} {2009})},\ \Eprint
  {http://arxiv.org/abs/0904.1537} {arXiv:0904.1537 [hep-ph]} \BibitemShut
  {NoStop}%
\bibitem [{\citenamefont {Germani}\ and\ \citenamefont {Kehagias}(2010)}]{new}%
  \BibitemOpen
  \bibfield  {author} {\bibinfo {author} {\bibfnamefont {C.}~\bibnamefont
  {Germani}}\ and\ \bibinfo {author} {\bibfnamefont {A.}~\bibnamefont
  {Kehagias}},\ }\href {\doibase 10.1103/PhysRevLett.105.011302} {\bibfield
  {journal} {\bibinfo  {journal} {Phys. Rev. Lett.}\ }\textbf {\bibinfo
  {volume} {105}},\ \bibinfo {pages} {011302} (\bibinfo {year}
  {2010})}\BibitemShut {NoStop}%
\bibitem [{\citenamefont {Di~Vita}\ and\ \citenamefont
  {Germani}(2016)}]{loop-new-higgs}%
  \BibitemOpen
  \bibfield  {author} {\bibinfo {author} {\bibfnamefont {S.}~\bibnamefont
  {Di~Vita}}\ and\ \bibinfo {author} {\bibfnamefont {C.}~\bibnamefont
  {Germani}},\ }\href {\doibase 10.1103/PhysRevD.93.045005} {\bibfield
  {journal} {\bibinfo  {journal} {Phys. Rev. D}\ }\textbf {\bibinfo {volume}
  {93}},\ \bibinfo {pages} {045005} (\bibinfo {year} {2016})}\BibitemShut
  {NoStop}%
\bibitem [{\citenamefont {Kehagias}\ \emph {et~al.}(2014)\citenamefont
  {Kehagias}, \citenamefont {Moradinezhad~Dizgah},\ and\ \citenamefont
  {Riotto}}]{kehagias-unitarity}%
  \BibitemOpen
  \bibfield  {author} {\bibinfo {author} {\bibfnamefont {A.}~\bibnamefont
  {Kehagias}}, \bibinfo {author} {\bibfnamefont {A.}~\bibnamefont
  {Moradinezhad~Dizgah}}, \ and\ \bibinfo {author} {\bibfnamefont
  {A.}~\bibnamefont {Riotto}},\ }\href {\doibase 10.1103/PhysRevD.89.043527}
  {\bibfield  {journal} {\bibinfo  {journal} {Phys. Rev. D}\ }\textbf {\bibinfo
  {volume} {89}},\ \bibinfo {pages} {043527} (\bibinfo {year}
  {2014})}\BibitemShut {NoStop}%
\bibitem [{\citenamefont {Xianyu}\ \emph {et~al.}(2013)\citenamefont {Xianyu},
  \citenamefont {Ren},\ and\ \citenamefont {He}}]{he1}%
  \BibitemOpen
  \bibfield  {author} {\bibinfo {author} {\bibfnamefont {Z.-Z.}\ \bibnamefont
  {Xianyu}}, \bibinfo {author} {\bibfnamefont {J.}~\bibnamefont {Ren}}, \ and\
  \bibinfo {author} {\bibfnamefont {H.-J.}\ \bibnamefont {He}},\ }\href
  {\doibase 10.1103/PhysRevD.88.096013} {\bibfield  {journal} {\bibinfo
  {journal} {Phys. Rev.}\ }\textbf {\bibinfo {volume} {D88}},\ \bibinfo {pages}
  {096013} (\bibinfo {year} {2013})},\ \Eprint {http://arxiv.org/abs/1305.0251}
  {arXiv:1305.0251 [hep-ph]} \BibitemShut {NoStop}%
\bibitem [{\citenamefont {Ren}\ \emph {et~al.}(2014)\citenamefont {Ren},
  \citenamefont {Xianyu},\ and\ \citenamefont {He}}]{he2}%
  \BibitemOpen
  \bibfield  {author} {\bibinfo {author} {\bibfnamefont {J.}~\bibnamefont
  {Ren}}, \bibinfo {author} {\bibfnamefont {Z.-Z.}\ \bibnamefont {Xianyu}}, \
  and\ \bibinfo {author} {\bibfnamefont {H.-J.}\ \bibnamefont {He}},\ }\href
  {\doibase 10.1088/1475-7516/2014/06/032} {\bibfield  {journal} {\bibinfo
  {journal} {JCAP}\ }\textbf {\bibinfo {volume} {1406}},\ \bibinfo {pages}
  {032} (\bibinfo {year} {2014})},\ \Eprint {http://arxiv.org/abs/1404.4627}
  {arXiv:1404.4627 [gr-qc]} \BibitemShut {NoStop}%
\bibitem [{\citenamefont {Bezrukov}\ \emph {et~al.}(2011)\citenamefont
  {Bezrukov}, \citenamefont {Magnin}, \citenamefont {Shaposhnikov},\ and\
  \citenamefont {Sibiryakov}}]{bezsib}%
  \BibitemOpen
  \bibfield  {author} {\bibinfo {author} {\bibfnamefont {F.}~\bibnamefont
  {Bezrukov}}, \bibinfo {author} {\bibfnamefont {A.}~\bibnamefont {Magnin}},
  \bibinfo {author} {\bibfnamefont {M.}~\bibnamefont {Shaposhnikov}}, \ and\
  \bibinfo {author} {\bibfnamefont {S.}~\bibnamefont {Sibiryakov}},\ }\href
  {\doibase 10.1007/JHEP01(2011)016} {\bibfield  {journal} {\bibinfo  {journal}
  {Journal of High Energy Physics}\ }\textbf {\bibinfo {volume} {2011}},\
  \bibinfo {pages} {16} (\bibinfo {year} {2011})}\BibitemShut {NoStop}%
\bibitem [{\citenamefont {Dicus}\ and\ \citenamefont
  {He}(2005{\natexlab{a}})}]{factorial}%
  \BibitemOpen
  \bibfield  {author} {\bibinfo {author} {\bibfnamefont {D.~A.}\ \bibnamefont
  {Dicus}}\ and\ \bibinfo {author} {\bibfnamefont {H.-J.}\ \bibnamefont {He}},\
  }\href {\doibase 10.1103/PhysRevD.71.093009} {\bibfield  {journal} {\bibinfo
  {journal} {Phys. Rev. D}\ }\textbf {\bibinfo {volume} {71}},\ \bibinfo
  {pages} {093009} (\bibinfo {year} {2005}{\natexlab{a}})}\BibitemShut
  {NoStop}%
\bibitem [{\citenamefont {Dicus}\ and\ \citenamefont
  {He}(2005{\natexlab{b}})}]{factor2}%
  \BibitemOpen
  \bibfield  {author} {\bibinfo {author} {\bibfnamefont {D.~A.}\ \bibnamefont
  {Dicus}}\ and\ \bibinfo {author} {\bibfnamefont {H.-J.}\ \bibnamefont {He}},\
  }\href {\doibase 10.1103/PhysRevLett.94.221802} {\bibfield  {journal}
  {\bibinfo  {journal} {Phys. Rev. Lett.}\ }\textbf {\bibinfo {volume} {94}},\
  \bibinfo {pages} {221802} (\bibinfo {year} {2005}{\natexlab{b}})},\ \Eprint
  {http://arxiv.org/abs/hep-ph/0502178} {arXiv:hep-ph/0502178 [hep-ph]}
  \BibitemShut {NoStop}%
\bibitem [{\citenamefont {Ferrara}\ \emph {et~al.}(2011)\citenamefont
  {Ferrara}, \citenamefont {Kallosh}, \citenamefont {Linde}, \citenamefont
  {Marrani},\ and\ \citenamefont {Van~Proeyen}}]{linde}%
  \BibitemOpen
  \bibfield  {author} {\bibinfo {author} {\bibfnamefont {S.}~\bibnamefont
  {Ferrara}}, \bibinfo {author} {\bibfnamefont {R.}~\bibnamefont {Kallosh}},
  \bibinfo {author} {\bibfnamefont {A.}~\bibnamefont {Linde}}, \bibinfo
  {author} {\bibfnamefont {A.}~\bibnamefont {Marrani}}, \ and\ \bibinfo
  {author} {\bibfnamefont {A.}~\bibnamefont {Van~Proeyen}},\ }\href {\doibase
  10.1103/PhysRevD.83.025008} {\bibfield  {journal} {\bibinfo  {journal} {Phys.
  Rev.}\ }\textbf {\bibinfo {volume} {D83}},\ \bibinfo {pages} {025008}
  (\bibinfo {year} {2011})},\ \Eprint {http://arxiv.org/abs/1008.2942}
  {arXiv:1008.2942 [hep-th]} \BibitemShut {NoStop}%
\bibitem [{\citenamefont {Bezrukov}\ \emph {et~al.}(2009)\citenamefont
  {Bezrukov}, \citenamefont {Gorbunov},\ and\ \citenamefont
  {Shaposhnikov}}]{higgsspectra}%
  \BibitemOpen
  \bibfield  {author} {\bibinfo {author} {\bibfnamefont {F.}~\bibnamefont
  {Bezrukov}}, \bibinfo {author} {\bibfnamefont {D.}~\bibnamefont {Gorbunov}},
  \ and\ \bibinfo {author} {\bibfnamefont {M.}~\bibnamefont {Shaposhnikov}},\
  }\href {\doibase 10.1088/1475-7516/2009/06/029} {\bibfield  {journal}
  {\bibinfo  {journal} {JCAP}\ }\textbf {\bibinfo {volume} {0906}},\ \bibinfo
  {pages} {029} (\bibinfo {year} {2009})},\ \Eprint
  {http://arxiv.org/abs/0812.3622} {arXiv:0812.3622 [hep-ph]} \BibitemShut
  {NoStop}%
\bibitem [{\citenamefont {Ade}\ \emph {et~al.}(2016)\citenamefont {Ade} \emph
  {et~al.}}]{planck}%
  \BibitemOpen
  \bibfield  {author} {\bibinfo {author} {\bibfnamefont {P.~A.~R.}\
  \bibnamefont {Ade}} \emph {et~al.} (\bibinfo {collaboration} {Planck}),\
  }\href {\doibase 10.1051/0004-6361/201525830} {\bibfield  {journal} {\bibinfo
   {journal} {Astron. Astrophys.}\ }\textbf {\bibinfo {volume} {594}},\
  \bibinfo {pages} {A13} (\bibinfo {year} {2016})},\ \Eprint
  {http://arxiv.org/abs/1502.01589} {arXiv:1502.01589 [astro-ph.CO]}
  \BibitemShut {NoStop}%
\bibitem [{\citenamefont {Germani}\ \emph {et~al.}(2016)\citenamefont
  {Germani}, \citenamefont {Kudryashova},\ and\ \citenamefont
  {Watanabe}}]{nina}%
  \BibitemOpen
  \bibfield  {author} {\bibinfo {author} {\bibfnamefont {C.}~\bibnamefont
  {Germani}}, \bibinfo {author} {\bibfnamefont {N.}~\bibnamefont
  {Kudryashova}}, \ and\ \bibinfo {author} {\bibfnamefont {Y.}~\bibnamefont
  {Watanabe}},\ }\href {\doibase 10.1088/1475-7516/2016/08/015} {\bibfield
  {journal} {\bibinfo  {journal} {JCAP}\ }\textbf {\bibinfo {volume} {1608}},\
  \bibinfo {pages} {015} (\bibinfo {year} {2016})},\ \Eprint
  {http://arxiv.org/abs/1512.06344} {arXiv:1512.06344 [astro-ph.CO]}
  \BibitemShut {NoStop}%
\bibitem [{\citenamefont {Germani}\ and\ \citenamefont
  {Watanabe}(2011)}]{yuki}%
  \BibitemOpen
  \bibfield  {author} {\bibinfo {author} {\bibfnamefont {C.}~\bibnamefont
  {Germani}}\ and\ \bibinfo {author} {\bibfnamefont {Y.}~\bibnamefont
  {Watanabe}},\ }\href {\doibase 10.1088/1475-7516/2011/07/031,
  10.1088/1475-7516/2011/07/A01} {\bibfield  {journal} {\bibinfo  {journal}
  {JCAP}\ }\textbf {\bibinfo {volume} {1107}},\ \bibinfo {pages} {031}
  (\bibinfo {year} {2011})},\ \bibinfo {note} {[Addendum:
  JCAP1107,A01(2011)]},\ \Eprint {http://arxiv.org/abs/1106.0502}
  {arXiv:1106.0502 [astro-ph.CO]} \BibitemShut {NoStop}%
\bibitem [{\citenamefont {Germani}\ \emph {et~al.}(2012)\citenamefont
  {Germani}, \citenamefont {Martucci},\ and\ \citenamefont
  {Moyassari}}]{parvin}%
  \BibitemOpen
  \bibfield  {author} {\bibinfo {author} {\bibfnamefont {C.}~\bibnamefont
  {Germani}}, \bibinfo {author} {\bibfnamefont {L.}~\bibnamefont {Martucci}}, \
  and\ \bibinfo {author} {\bibfnamefont {P.}~\bibnamefont {Moyassari}},\ }\href
  {\doibase 10.1103/PhysRevD.85.103501} {\bibfield  {journal} {\bibinfo
  {journal} {Phys. Rev.}\ }\textbf {\bibinfo {volume} {D85}},\ \bibinfo {pages}
  {103501} (\bibinfo {year} {2012})},\ \Eprint {http://arxiv.org/abs/1108.1406}
  {arXiv:1108.1406 [hep-th]} \BibitemShut {NoStop}%
\bibitem [{\citenamefont {Germani}\ \emph {et~al.}(2014)\citenamefont
  {Germani}, \citenamefont {Watanabe},\ and\ \citenamefont
  {Wintergerst}}]{newhiggsspectra}%
  \BibitemOpen
  \bibfield  {author} {\bibinfo {author} {\bibfnamefont {C.}~\bibnamefont
  {Germani}}, \bibinfo {author} {\bibfnamefont {Y.}~\bibnamefont {Watanabe}}, \
  and\ \bibinfo {author} {\bibfnamefont {N.}~\bibnamefont {Wintergerst}},\
  }\href {\doibase 10.1088/1475-7516/2014/12/009} {\bibfield  {journal}
  {\bibinfo  {journal} {JCAP}\ }\textbf {\bibinfo {volume} {1412}},\ \bibinfo
  {pages} {009} (\bibinfo {year} {2014})},\ \Eprint
  {http://arxiv.org/abs/1403.5766} {arXiv:1403.5766 [hep-ph]} \BibitemShut
  {NoStop}%
\bibitem [{\citenamefont {Barbon}\ and\ \citenamefont
  {Espinosa}(2009)}]{espinosa}%
  \BibitemOpen
  \bibfield  {author} {\bibinfo {author} {\bibfnamefont {J.~L.~F.}\
  \bibnamefont {Barbon}}\ and\ \bibinfo {author} {\bibfnamefont {J.~R.}\
  \bibnamefont {Espinosa}},\ }\href {\doibase 10.1103/PhysRevD.79.081302}
  {\bibfield  {journal} {\bibinfo  {journal} {Phys. Rev.}\ }\textbf {\bibinfo
  {volume} {D79}},\ \bibinfo {pages} {081302} (\bibinfo {year} {2009})},\
  \Eprint {http://arxiv.org/abs/0903.0355} {arXiv:0903.0355 [hep-ph]}
  \BibitemShut {NoStop}%
\bibitem [{\citenamefont {Burgess}\ \emph {et~al.}(2010)\citenamefont
  {Burgess}, \citenamefont {Lee},\ and\ \citenamefont {Trott}}]{burgess}%
  \BibitemOpen
  \bibfield  {author} {\bibinfo {author} {\bibfnamefont {C.~P.}\ \bibnamefont
  {Burgess}}, \bibinfo {author} {\bibfnamefont {H.~M.}\ \bibnamefont {Lee}}, \
  and\ \bibinfo {author} {\bibfnamefont {M.}~\bibnamefont {Trott}},\ }\href
  {\doibase 10.1007/JHEP07(2010)007} {\bibfield  {journal} {\bibinfo  {journal}
  {JHEP}\ }\textbf {\bibinfo {volume} {07}},\ \bibinfo {pages} {007} (\bibinfo
  {year} {2010})},\ \Eprint {http://arxiv.org/abs/1002.2730} {arXiv:1002.2730
  [hep-ph]} \BibitemShut {NoStop}%
\bibitem [{\citenamefont {Hertzberg}(2010)}]{hertzberg}%
  \BibitemOpen
  \bibfield  {author} {\bibinfo {author} {\bibfnamefont {M.~P.}\ \bibnamefont
  {Hertzberg}},\ }\href {\doibase 10.1007/JHEP11(2010)023} {\bibfield
  {journal} {\bibinfo  {journal} {JHEP}\ }\textbf {\bibinfo {volume} {11}},\
  \bibinfo {pages} {023} (\bibinfo {year} {2010})},\ \Eprint
  {http://arxiv.org/abs/1002.2995} {arXiv:1002.2995 [hep-ph]} \BibitemShut
  {NoStop}%
\bibitem [{\citenamefont {Bezrukov}\ and\ \citenamefont
  {Shaposhnikov}(2008)}]{shaposhnikov}%
  \BibitemOpen
  \bibfield  {author} {\bibinfo {author} {\bibfnamefont {F.~L.}\ \bibnamefont
  {Bezrukov}}\ and\ \bibinfo {author} {\bibfnamefont {M.}~\bibnamefont
  {Shaposhnikov}},\ }\href {\doibase 10.1016/j.physletb.2007.11.072} {\bibfield
   {journal} {\bibinfo  {journal} {Phys. Lett.}\ }\textbf {\bibinfo {volume}
  {B659}},\ \bibinfo {pages} {703} (\bibinfo {year} {2008})},\ \Eprint
  {http://arxiv.org/abs/0710.3755} {arXiv:0710.3755 [hep-th]} \BibitemShut
  {NoStop}%
\bibitem [{\citenamefont {Calmet}\ and\ \citenamefont
  {Casadio}(2014)}]{casadio}%
  \BibitemOpen
  \bibfield  {author} {\bibinfo {author} {\bibfnamefont {X.}~\bibnamefont
  {Calmet}}\ and\ \bibinfo {author} {\bibfnamefont {R.}~\bibnamefont
  {Casadio}},\ }\href {\doibase 10.1016/j.physletb.2014.05.008} {\bibfield
  {journal} {\bibinfo  {journal} {Phys. Lett.}\ }\textbf {\bibinfo {volume}
  {B734}},\ \bibinfo {pages} {17} (\bibinfo {year} {2014})},\ \Eprint
  {http://arxiv.org/abs/1310.7410} {arXiv:1310.7410 [hep-ph]} \BibitemShut
  {NoStop}%
\end{thebibliography}%

\end{document}